\documentclass[12pt]{iopart}
\usepackage{graphics}
\usepackage{epsfig}
\usepackage{psfrag}

\begin{document}

\newcommand{\gtrsim}{\,\rlap{\lower 3.5 pt \hbox{$\mathchar \sim$}} \raise
1pt \hbox {$>$}\,}
\newcommand{\lesssim}{\,\rlap{\lower 3.5 pt \hbox{$\mathchar \sim$}} \raise
1pt \hbox {$<$}\,}

\title{Observational constraints on particle production during inflation}
\author{{\O}ystein Elgar{\o}y}
\address{NORDITA, Blegdamsvej 17, DK-2100 Copenhagen, Denmark}
\author{Steen Hannestad}
\address{Department of Physics, University of Southern Denmark, 
Campusvej 55, DK-5230 Odense M, Denmark \\
and
\\
NORDITA, Blegdamsvej 17, DK-2100 Copenhagen, Denmark}
\author{Troels Haugb{\o}lle}
\address{Department of Astronomy, Niels Bohr Institute, \\
Juliane Maries Vej 30, DK-2100 Copenhagen, Denmark}

\date{{\today}}

\begin{abstract}
Resonant particle production, along with many other physical
processes which change the effective equation of state (EOS) during
inflation, introduces a feature in the primordial power
spectrum which in many models has a step-like shape.
We calculate observational constraints on resonant particle production,
parametrized in the form of an effective step height, $N_{\rm eff}$
and location in $k$-space, $k_{\rm break}$. 
Combining data from the cosmic microwave
background and the 2dF Galaxy Redshift Survey yields strong constraints in
some regions of parameter space, although the range in $k$-space which
can be probed is restricted to 
$k \sim 0.001 - 0.1h~{\rm Mpc}^{-1}$. 
We also discuss the implications of our findings for general models
which change the effective EOS during inflation.
\end{abstract}
\maketitle

\section{Introduction}

The recent measurements of the temperature anisotropies in the 
cosmic microwave background (CMB) 
from the Wilkinson Microwave Anisotropy Probe (WMAP) 
\cite{map1}--\cite{map5} 
have set a new standard for precision in cosmology.  In addition to pinning 
down several key cosmological parameters with unprecedented accuracy, 
the WMAP data have confirmed important aspects of the inflationary 
paradigm.  Quantum fluctuations during the inflationary phase are commonly 
thought to be the seeds of the anisotropies in the CMB and large-scale 
structure in the Universe.   
The observations are consistent with the inflationary 
mechanism for generating curvature fluctuations on superhorizon 
scales, and with the fluctuations being adiabatic and Gaussian 
as in the simplest single-field models \cite{map6}.    
The fluctuations are 
conveniently characterized by the power spectrum of the curvature 
perturbations ${\mathcal P}_{\mathcal R}(k)$, and in the simplest models 
it is proportional to some power of the comoving wavenumber $k$ of 
a given Fourier mode, ${\mathcal P}_{\mathcal R}(k) \propto k^{n_s-1}$, 
where $n_s$ is the so-called scalar spectral index.  For $n_s=1$ 
one has the case of scale-invariant fluctuations.  
The data favour a slight red tilt of the scalar spectral index, 
$n_s < 1$, again consistent single-field models \cite{map2}.

However, there are several possible extensions to the simplest 
models for inflation, and in most of them deviations from the 
simple power-law form of the primordial power spectrum are 
produced.  Models which produce features in the primordial power spectrum 
include variants of extended inflation \cite{la}, 
models with multiple episodes 
of inflation \cite{sarkar}, 
and phase transitions during the inflationary phase \cite{phasetrans}.  
In this paper we consider another proposal, resonant particle 
production during inflation, which extends the single-field models 
by coupling the inflaton to a massive field (see e.g.~\cite{Chung:1999a}).  
By this mechanism 
energy can be extracted from the inflaton field during inflation, 
altering its classical motion and producing features in the primordial 
power spectrum.  

Our aim in this paper is to calculate in detail the 
shape of the primordial power spectrum in the presence of 
resonant particle production, 
and to constrain this class of models using a collection of the 
latest cosmological data, including the WMAP CMB power spectrum and 
the 2dF Galaxy Redshift Survey (2dFGRS) galaxy power spectrum 
\cite{percival}.    
It is important to bear in mind that ${\mathcal P}_{\mathcal R}(k)$ is 
not directly observable.  Whether we look at the CMB or the 
large-scale distribution of matter, what we see is the primordial 
fluctuations modified by astrophysical processes.  Thus, our ability 
to constrain the primordial fluctuations depends on our understanding 
of the physics involved in the processing, and on our knowledge of 
the relevant parameters controlling the physics, for instance the 
matter density.  Combining different cosmological observations 
is therefore essential, since different observations probe different 
combinations of cosmological parameters, and thus combining them 
breaks parameter degeneracies and allows tighter constraints to be 
obtained.  

The structure of this paper is as follows.  In section 2 we describe 
the model and the calculation of the primordial power spectrum in 
the presence of resonant particle production, and in section 3 we 
comment on the generality of our results.  The observations we use, 
CMB data and the galaxy power spectrum, are described in section 4, 
and section 5 discusses the likelihood analysis and the results, 
before we conclude in section 6.  

\section{Resonant particle production}

\looseness=-1
We will work with a simple single field model for the inflaton. 
We do this because, as will be seen, the main interest here is not
the slow change in the power spectrum during the course of inflation, but rather,
what happens if a resonant phenomenon---namely a sudden burst of particle 
production---radically changes the dynamics for a short period of time.

Our specific toy model is that of a single scalar field $\phi$, 
dubbed the inflaton, which is rolling slowly down its
potential in accordance with the slow-roll conditions, 
dominating the energy--momentum tensor.  
The inflaton is coupled through the
mass to a massive boson $\varphi$ in such a way that at a certain moment $t_0$
$\varphi$ becomes massless. Initially $\varphi$,  being heavy compared to the
natural mass scale set by the expansion rate $m^2_{natural} \simeq H^2$, 
is in its vacuum state. The potential is
\begin{equation}\label{eq:toymodel}
V(\phi,\varphi) = V(\phi) + \case{1}{2}{\mathcal N}(m - g\phi)^2\varphi^2  ,
\end{equation}
where we will allow for the case where $\varphi$ represents
some degenerate species and we denote the number of different species by ${\mathcal N}$ 
as done in \cite{Chung:1999a}.
The phenomenon is strongly resonant essentially because any
particles created will be red shifted away exponentially fast.

To find the backreaction we will treat the inflaton $\phi$ at the classical
level, while the $\varphi$ field is fully quantized.  Computing
the backreaction implies finding the expectation value of the
energy--momentum tensor associated with the $\varphi$ field. This is,
due to problems with proper regularization and renormalization, a
highly nontrivial problem (see e.g.~\cite{Kofman:1997a}--\cite{Cormier:2001a}
and \cite{baacke:1998a, baacke:1999a} for the fermionic case), but as a lowest
approximation we can use the Hartree approximation.

\subsection{Formalism and equations of motion}
We assume a spatially flat Friedmann--Robertson--Walker (FRW) model overlayed with linear
scalar perturbations and the metric takes the form \cite{Mukhanov:1992a}
\begin{equation}\label{eq:metric}
\fl\rmd s^2 = -(1 + 2A)\rmd t^2 + 2a(t)\partial_iB \rmd x^i \rmd t 
+ a^2(t) [(1-2\psi)\delta_{ij} + 2\partial_i\partial_jE]\rmd x^i\rmd x^j,
\end{equation}
where 
$a(t)$ is the scale factor and $A$, $B$, $\psi$ and $E$ describe the possible
scalar metric perturbations. Likewise given $N$ scalar fields $\chi^I$, $I=1,\ldots,N$, 
they can be split into a homogeneous part and a fluctuation
\begin{equation}\label{eq:fluctdef}
\chi^I(t,x) = \chi^I(t) + \delta\chi^I(t,x).
\end{equation}

Because of the inherent gauge freedom related to the arbitrariness of coordinates, we
can eliminate two out of the four metric perturbations by choosing the spatial and
temporal hypersurfaces in a proper way \cite{Abramo:1997b, Bruni:1997a}.
The physical content is of course not changed by such a 
coordinate---or gauge---transformation. Certain algebraic combinations of the metric and matter
perturbations in {\it any} gauge can be found that correspond to a certain
perturbation in a {\it specific gauge}. These combinations are dubbed {\it gauge invariant
variables}.

In this paper we will use the spatially flat gauge to
describe the field perturbations. These are called the Sasaki--Mukhanov variables:
\begin{equation}
Q^I = \delta\chi^I + \frac{\dot\chi^I}{H}\psi.
\end{equation}
The fourier transform $Q^I_k$ of $Q^I$ has the evolution equation \cite{Gordon:2000a}
\begin{equation}\label{eq:evolQ}
\ddot{Q}^I_k + 3H\dot{Q}^I_k +\frac{k^2}{a^2}Q^I_k
                      + \sum_J {\mathcal M}^I_J Q^J_k = 0  ,
\end{equation}
where ${\mathcal M}$ is the effective mass matrix
\begin{equation}\label{eq:massmatrix}
{\mathcal M}^I_{\phantom{I}J} = \frac{\partial^2 V}{\partial\chi^I\partial\chi^J}
                   - \frac{8\pi}{m^2_{pl}a^3}
                   \left(\frac{a^3\dot{\chi}^I\dot{\chi}^J}{H}\right)^\cdot.
\end{equation}

A popular measure of the curvature perturbation, which later directly relates to the
temperature perturbation in the CMBR, is ${\mathcal R}$---the spatial curvature seen by observers
who are comoving with the energy content. It is conveniently related to the
Sasaki--Mukhanov variables as \cite{Wands:2000a}
\begin{equation}\label{eq:R}
{\mathcal R} = H \times \sum_I \left[ \left( 
               \frac{\dot{\chi}^I}{\sum_J \dot{\chi}^J\dot{\chi}^J}
\right) Q_I \right].
\end{equation}
In our case $\{\chi^I\}=\{\phi,\varphi\}$.

The evolution of the background parameters, taking into account the backreaction of the
$\varphi$ field, but not rescattering of the inflation field itself is found to 
be:
\begin{eqnarray}
\label{eq:eq1}
\ddot{\phi}  =  -3H\dot{\phi} - \frac{\rmd V(\phi)}{\rmd\phi} +
                 g{\mathcal N} (m-g\phi)\langle\varphi^2\rangle, \\
H^2  =  \frac{4\pi}{3m_{pl}^2} \left[ \dot{\phi}^2 + 2V(\phi) +
    {\mathcal N}\langle\dot{\varphi}^2\rangle + 
\label{eq:eq4}  {\mathcal N}(m-g\phi)^2
    \langle\varphi^2\rangle+
    \frac{1}{a^2}\langle\nabla\varphi^2\rangle \right], \\ 
\label{eq:eq5}
\dot{H}  =  -\frac{4\pi}{m_{pl}^2}\left[ \dot{\phi}^2 + \langle\dot{\varphi}^2\rangle +
    \frac{1}{3a^2}\langle\nabla\varphi^2\rangle \right].
\end{eqnarray}
The expectation values $\langle\cdot\rangle$ will be computed below.

\subsection{Quantization of the $\varphi$ field}
When we quantize the $\varphi$ field we will disregard any effects 
arising from the metric perturbations. For example, we will consider $\delta\varphi$
instead of $Q^\varphi_k$ effectively evolving according to equation~(\ref{eq:evolQ})
without the second term in equation~(\ref{eq:massmatrix}). This is reasonable as long as 
we are looking at wavelengths shorter than the horizon size. We will
see below that indeed the bulk of the energy is deposited in sub-horizon
modes justifying the approach. The backreaction of $\phi$-modes on the
$\varphi$-field can also be disregarded, because it is an almost homogeneous
light field.

When quantizing $\varphi$ we have chosen to use the rescaled field $a^{3/2}\varphi$.
This is done in order to get rid of trivial prefactors arising from the
expansion of background spacetime. Working in the Heisenberg picture the $\varphi$ field 
can be canonically quantized and decomposed in terms of annihilation and
creation operators
\begin{equation}\label{eq:quantization}
a^{3/2}\hat{\varphi} = \frac{1}{(2\pi)^{3/2}}\int
\rmd^3k\hat{a}_k\left(a^{3/2}\varphi_k\right)(t)\e^{-\rmi kx}
+\hat{a}^\dagger_k\left(a^{3/2}\varphi_k\right)^*(t)\e^{+\rmi kx},
\end{equation}
satisfying the usual commutation relations
\begin{equation}\label{eq:commun}
[\hat{a}_k, \hat{a}_{k'}] = [\hat{a}_k^\dagger, \hat{a}_{k'}^\dagger]
= 0, \qquad [\hat{a}_k, \hat{a}_{k'}^\dagger] = \delta^3(\vec{k}-\vec{k'}).
\end{equation}
By construction $\varphi_k$ has to be a solution to the classical Klein--Gordon 
equation of a massive field with time-dependent mass
\begin{equation}\label{eq:psik}
\left(a^{3/2}\varphi_k\right)\ddot{\phantom{)}} + \omega_k^2 \left(a^{3/2}\varphi_k\right) = 0  ,
\end{equation}
\begin{equation}\label{eq:psik2}
\omega_k^2 = \frac{k^2}{a^2} + (m-g\phi)^2 - \frac{9}{4}H^2-\frac{3}{4}\dot{H} .
\end{equation}
Except for the very small time interval in which the particles are produced,
the field is heavy compared to the `Hubble mass'
\begin{equation}
(m-g\phi)^2\gg\case{3}{2}H^2,
\end{equation}
and the vacuum fluctuations are exponentially damped on large scales. When
$(m-g\phi)^2<\frac{3}{2}H^2$ particle production occurs, and we can assume it to
be the dominant effect. Therefore we will disregard vacuum polarization
altogether. To understand the qualitative behaviour of the dynamics, we do not
need to include the terms coming from the metric equation~(\ref{eq:massmatrix}) nor terms in
equation~(\ref{eq:psik2}) that are dependent on the Hubble parameter.

Initially $\varphi_k$ is in the vacuum state of a heavy bosonic field
\begin{equation}\label{eq:massive}
a^{3/2}\varphi_k = \frac{1}{\sqrt{2\omega_k}}\exp\left(-\rmi\int^t\omega_k \rmd t\right),
\end{equation}
while after some moment $t_0$, where
\begin{equation}
m - g\phi_0 = 0  ,
\end{equation}
particle production, because of interaction with the inflaton, has
occurred. Generally $\varphi_k$ can be written in terms of `adiabatic' solutions
to equation~(\ref{eq:psik})
\begin{equation}\label{eq:bog}
a^{3/2}\varphi_k(t) = \frac{\alpha_k}{\sqrt{2\omega_k}}\exp\left(-\rmi\int^t\omega_k \rmd t\right) +
             \frac{\beta_k}{\sqrt{2\omega_k}}\exp\left(+\rmi\int^t\omega_k \rmd t\right).
\end{equation}
Then initially $\alpha_k=1$, $\beta_k=0$, while at later times the effective mass $(m-g\phi)$
only changes slowly and
equation~(\ref{eq:bog}) is a solution to equation~(\ref{eq:psik}) {\it with constant} $\alpha_k$,
$\beta_k$.

\subsection{Characteristics of the $\varphi$ field}
Choosing a special boundary condition for equation~(\ref{eq:psik}) is
equivalent to singling out a special set of observers (and
annihilation and creation operators). We have already mentioned
one, namely equation~(\ref{eq:massive}), which is related to the vacuum seen
by observers in the infinite past. Another interesting vacuum is
the comoving vacuum related to the observers which are comoving at
a given time.

Different vacua are related to each other by a Bogoliubov
transformation\footnote[4]{Strictly speaking this is only true for vacua which
share the same spatial base \cite{birrell}.}
\begin{equation}
\tilde\varphi_k(t) = \alpha_k\varphi_k(t) + \beta_k\varphi_k(t).
\end{equation}
We can from this transformation directly calculate distribution functions and
expectation values of a field in the $\left| 0 \right\rangle$ vacuum state, seen by an
observer corresponding to the $\left| \tilde0 \right\rangle$ vacuum state. Notice that
as a consequence of our parametrization equation~(\ref{eq:bog}), we can read off the Bogoliubov
coefficients for a transformation from the initial vacuum state to a comoving state to be
$\alpha_k, \beta_k$ as defined in equation~(\ref{eq:bog}).

If we suppose that the particle production is resonant around $t_0$,
the effective mass of $\varphi$ can be Taylor expanded near $t_0$.
Furthermore if everything happens approximately during one e-fold
we will also suppose $a\simeq a_0$ and the equation of motion for
the $\varphi$-modes equation~(\ref{eq:psik}) in this neighbourhood is
\begin{equation}
\left(a^{3/2}\varphi_k\right)\ddot{\phantom{)}} + \left(\frac{k^2}{a_0^2} + g^2|\dot{\phi}_0|^2(t-t_0)^2\right)\left(a^{3/2}\varphi_k\right) = 0 .
\end{equation}
Now we can rewrite equation~(\ref{eq:psik}) in terms of the dimensionless momentum $p^2\equiv\sigma^{-1}k^2/a_0^2$
and the time $T^2=\sigma (t-t_0)^2$, where $\sigma = g|\phi_0|$,
\begin{equation}
\frac{\rmd^2}{\rmd T^2}\left(a^{3/2}\varphi_p\right) + \left(p^2 + T^2\right)\left(a^{3/2}\varphi_p\right) = 0.
\end{equation}
This is equivalent to a Schr\"odinger equation for scattering at a (negative) parabolic potential and can be
solved in terms of parabolic cylinder functions. The incoming wave can be approximated as pure positive
frequency corresponding to being in the vacuum state, while the outgoing state $(\alpha_k, \beta_k)$ is
found to be (see \cite{Kofman:1997a} for details)
\begin{eqnarray}
\label{eq:bogocoef}
\alpha_k^{out} = \sqrt{1 + \e^{-\pi p^2}}\e^{\rmi\zeta_k}  , \tqs
\beta_k^{out}  = -\rmi \e^{-\frac{1}{2}\pi p^2 - 2\rmi\zeta_k}, \\
\zeta_k        = \arg\Gamma\left(\frac{1+\rmi p^2}{2}\right) + \frac{p^2}{2}\left(1-\ln \frac{p^2}{2}\right).
\end{eqnarray}
The number density of $\varphi_k$ modes after particle production is given as \cite{birrell}
\begin{equation}
\fl a^3 n_k(t)  =  \left\langle 0 \left| \hat N_k \right| 0 \right\rangle =
                       \left\langle 0 \left| \hat a^\dagger_k(t) \hat a_k(t) \right| 0 \right\rangle  
          =  |\beta_k^{out}|^2 = \e^{-\pi p^2} = \e^{-\pi k^2/g|\dot{\phi}_0|a_0^2}. \label{eq:numberdens}
\end{equation}
If we suppose that the typical time scale of the inflaton field is
$H^{-1}$ then $\dot{\phi}_0\simeq\phi_0H_0$. From the slow-roll
result and the limit on the quadrupole we find that if particle
production has happened at an observationally interesting scale,
the exponent above becomes 
\[-\frac{\pi
H_0}{m}\left(\frac{k/a_0}{H_0}\right)^2.
\]
 This goes to show that
initially the spectrum of $\varphi$-modes will be quite blue.
Integrating equation~(\ref{eq:numberdens}) we find the total number density to
be
\begin{equation}\label{eq:totaldens}
n_\varphi(t) = \frac{1}{(2\pi)^3}\int \rmd k^3 n_k = g^{3/2}\frac{|\dot{\phi}_0|^{3/2}}{(2\pi)^3}\left(\frac{a}{a_0}\right)^3
                   \equiv n_0 \left(\frac{a}{a_0}\right)^3.
\end{equation}
When we try to calculate the two-point correlator
\mbox{$\langle\varphi^2\rangle$} ultraviolet divergences arise.
These can be dealt with exactly like in normal flat Minkowski-space by subtracting the vacuum contribution \cite{Chung:1999a}
and we find
\begin{eqnarray}
\fl\langle\varphi^2\rangle = \frac{1}{2\pi^2a^3}\int \frac{\rmd kk^2}{\omega_k}
\left[|\beta_k|^2
               + Re\left(\alpha_k\beta_k^*\exp\left(-2\rmi\int^t\omega_k\rmd t\right)\right)\right]\nonumber\\
        \lo\simeq \frac{{\mathcal C}}{m-g\phi}n_0\left( \frac{a}{a_0} \right)^{-3} ,\label{eq:psisquared}
\end{eqnarray}
where ${\mathcal C}=0.680$ is a numerical constant\footnote{${\mathcal C}=\int p^2\rmd p \e^{-\pi p^2}(1-\sqrt{1+\e^{\pi p^2}}\sin\zeta_k)
/ \int p^2\rmd p \e^{-\pi p^2}$. The correction ${\mathcal C}$ is due to the difference in expanding $|\beta_k|^2+Re(\ldots)$ correctly (nominator) compared to the approximation used (denominator).}.

\subsection{The inflaton field}
Energy from the created $\varphi$ particles is draining the kinetic energy of the
background inflaton field, but in turn the $\varphi$ field reacts back on the fluctuations
of the inflaton. The inflaton field is a light field, it is frozen on large scales, and it
is important to use the full equation of motion equation~(\ref{eq:evolQ}) including metric terms
for its fluctuations $Q_k$.

The homogeneous part of the $\varphi$ field is negligible to begin
with, because the field is overdamped on large scales. If this is
the case, and we later disregard oscillations between the two
fields, i.e.~correlators of the form $\langle\phi\varphi\rangle$,
then the homogeneous part of the $\varphi$ field is easily seen to
remain negligible. A similar analysis of second-order effects from
a scalar field coupled to the inflaton during preheating
\cite{Liddle:99} found no impact from linear perturbations, and indeed
showed that the dominant effect is second order in the scalar field
perturbation. This in turn means that we can disregard any
off-diagonal terms in the equation of motion for the perturbations
of the $\phi$ field. Their equation of motion, using
equation~(\ref{eq:evolQ}), is then
\begin{eqnarray}\nonumber
\fl\ddot{Q}_k   =  - 3H\dot{Q}_k - \Bigg[\frac{k^2}{a^2} + 
\frac{\rmd^2V(\phi)}{\rmd\phi^2} + g^2{\mathcal N}\langle\varphi^2\rangle \\
\label{eq:evolQ2}
   +\frac{8\pi}{m^2_{pl}H} \bigg(\bigg(3H + \frac{\dot{H}}{H}\bigg)\dot{\phi} 
+ 2\frac{\rmd V(\phi)}{\rmd\phi}
- g{\mathcal N}(m-g\phi)\langle\varphi^2\rangle\bigg)\Bigg]Q_k ,
\end{eqnarray}
where we naively have replaced any quadratic terms in $\varphi$ with the corresponding correlators 
and $Q_k=\delta\phi + (\dot{\phi}/H)\psi$.

In a proper treatment one should include other second-order terms in
equations~(\ref{eq:evolQ2}), (\ref{eq:eq1}), (\ref{eq:eq4}) and
(\ref{eq:eq5}) coming from metric perturbations
\cite{Abramo:1997b,Bruni:1997a,Bartolo:2002,Maldacena:2002,Rigopoulos:2002}, but
most probably it will only lead to a quantitative change of the produced feature in the power spectrum,  not a qualitative one. 

\subsection{Numerical results}
In order to find the impact on the power spectrum, we carried out a
numerical integration of the whole system including the two-point
correlators such as the one in equation~(\ref{eq:psisquared}). The obvious
strategy would be to evaluate the two-point correlators in terms
of the Bogoliubov coefficients, but this scheme turns out to be numerically
unstable. Instead we use the idea of \cite{Cormier:2001a} and
parametrize $\varphi_k$ as
\begin{equation}\label{eq:fdef}
\varphi_k \equiv \frac{\exp\left(-\rmi\int^{t'}\omega_k\rmd t'\right)}{\sqrt{2\omega_k}a^{3/2}} [1+f].
\end{equation}
Making $f$ the dynamical variable we get the advantage that the
two-point correlators can be evaluated directly in terms of $f$.
After some algebra, we find all the two point correlators: 
\begin{eqnarray}
\label{eq:fvpsq}
\fl\langle \varphi^2 \rangle  =  \frac{1}{2\pi^2a^3}\int\frac{k^2\rmd k}{\omega_k}
         \left[ \frac{1}{2}|f|^2 + Re(f) \right], \\ \label{eq:fvpdsq}
\fl\langle \varphi\dot{\varphi} \rangle  =  \frac{1}{4\pi^2a^3}\int\frac{k^2\rmd k}{\omega_k}
      \bigg[ Re((1+f)\dot{f}) - \left(\frac{\dot{\omega}_k}{2\omega_k} +
   \frac{3}{2} H \right)(|f|^2 + 2 Re(f)) \bigg], \\ \nonumber
\fl\langle \dot{\varphi}^2 \rangle  =  \frac{1}{4\pi^2a^3}\int\frac{k^2\rmd k}{2\omega_k}
     \bigg[ |\dot{f}|^2 + \bigg(\Big(\frac{\dot{\omega}_k}{2\omega_k}
           + \frac{3}{2}H\Big)^2    + \omega_k^2 \bigg) 
   \left(|f|^2 + 2Re(f)\right) \\ \nonumber  + 
  2Re(\dot{f})\bigg( \omega_k Im(f) - \left( 1 + Re(f) \right) 
 \Big(\frac{\dot{\omega}_k}{2\omega_k} + \frac{3}{2}H\Big) \bigg) 
\\   + 2 Im(\dot{f}) \bigg(\omega_k  \left( 1 + Re(f) \right)
 + \Big(\frac{\dot{\omega}_k}{2\omega_k} + \frac{3}{2}H\Big)
           Im(f) \bigg) \bigg]. \label{eq:fvpnsq}
\end{eqnarray}
We could instead use $\varphi_k$ directly to compute
e.g.~$\langle \varphi^2 \rangle$, then 
\begin{equation}
\langle \varphi^2 \rangle = \frac{1}{2\pi^2}\int k^2\rmd k
           \left[ \left|\varphi_k \right|^2 - \frac{1}{2a^3\omega_k} \right].
\end{equation}
It was found numerically that evaluating the correlators directly in terms of
$\varphi_k$ is prone to numerical instability, because the above integral has
to be performed as a discreet sum over modes, where even small roundoff or numerical
error in $\varphi_k$ for high $k$ leads to large errors in the resulting correlator.
In effect, using $f$ instead, we only evolve the deviation from the vacuum mode.

Notice that in all of the above expressions the fictive
contribution from the vacuum expectation value has been subtracted, 
rendering them finite.

Inserting the definition of $f$ from equation~(\ref{eq:fdef}) into equation~(\ref{eq:psik}) we find its
evolution equation
\begin{equation}\label{eq:fevol}
\ddot{f} = \left(2i\omega_k + \frac{\dot{\omega}_k}{\omega_k}\right)\dot{f}
            - \bigg[ \frac{3}{4}\left(\frac{\dot{\omega}_k}{\omega_k}\right)^2
            - \frac{\ddot{\omega}_k}{2\omega_k} -\frac{9}{4}H^2 - \frac{3}{4}\dot{H}\bigg](1+f).
\end{equation}

To make any actual integration we have to choose  a specific inflationary model by specifying 
the inflaton potential $V(\phi)$.  For ease of comparison, 
we have chosen to use the same model as
Chung {\it et al\/}~\cite{Chung:1999a}: 
\begin{equation}
\label{eq:model1}
V(\phi) = \case{1}{2}m_\phi^2\phi^2   , \tqs m_\phi = 10^{-6}m_{pl}   , \tqs
m=2 m_{pl}   , \tqs g=1,
\end{equation}
with varying ${\mathcal N}$. To check if the slope of the slow-roll potential had any impact, several
runs with the exponential potential
\begin{equation}\eqalign{\label{eq:model2}
V(\phi)  =  \case{1}{2}m_\phi^2m_{pl}^2\exp\left(\alpha\frac{\phi}{m_{pl}}\right)   ,
\tqs  \alpha = 0.7034   ,  \cr
 m_\phi = 10^{-6}m_{pl}   , \tqs  m=2  m_{pl}   , \tqs g=1,}
\end{equation}
were carried out. The values of $\alpha$ and $m_\phi$ are chosen such
that $\dot{\phi}$ and the Hubble parameter $H$ are exactly the same as
in equation~(\ref{eq:model1}) at the moment of particle production. This
potential is ideally suited to test if the inflaton potential 
has any impact, because it has a
substantially higher acceleration of the inflaton. We found
that the effective step height $N_{\rm eff}$  is
virtually unaltered between the two models, confirming the
resonant nature of the phenomena.

By starting the simulation at a point so far back in time that all
modes of interest were way inside the horizon, we used the
Bunch--Davies vacuum state equation~(\ref{eq:massive}) as initial values for
$\varphi_k$ and $Q_k$. Furthermore the inflaton was taken to be
slow-rolling as an initial condition:
\begin{eqnarray}
 \phi(0)  =  \phi_{start},\tqs
\dot{\phi}(0) =  - \frac{V,_\phi(\phi_{start})}{3H(0)} ,\tqs
H^2(0) = \frac{8\pi}{3m_{pl}^2}V(\phi_{start}) ,\\
Q_k(0)  =  \frac{e(k)}{\sqrt{2k} a^{3/2}},\tqs
\dot{Q}_k(0) =  - \left(\frac{3}{2}H + \rmi \frac{k}{a_0}\right) Q_k(0), \\
 f  =  0 ,\tqs \dot{f} = 0   ,
\end{eqnarray}
where $e(k)$ is a random variable on the complex unit circle. To
check that these values had no impact on the further evolution, 
different starting values $\phi_{start}$ were chosen for different runs, and
in general $a_{start}$ was tuned such that at $m-g\phi_0=0$ we have $a_0=1$.

To get the proper attractor solution of the background inflaton
field, we let the simulation run for some time only evolving the homogeneous
component of the inflaton until the initial solution has relaxed to the 
correct attractor solution. Afterwards we start evolving the $\varphi$- and $\phi$-modes.
\begin{figure}[!htbp]
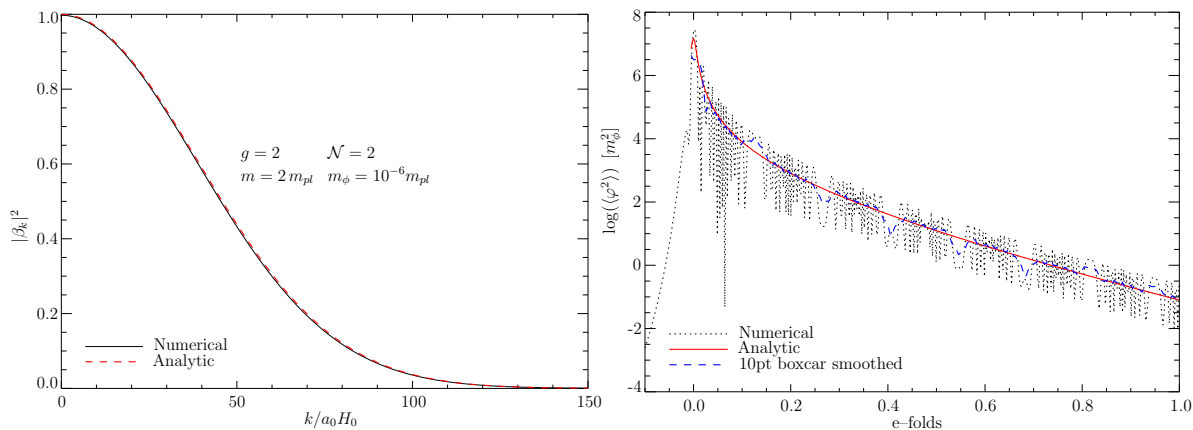

\centering
\begin{minipage}[b]{0.5\textwidth}
 \centering\includegraphics[width=\textwidth]{numberdens.epsi}
\end{minipage}%
\begin{minipage}[b]{0.5\textwidth}
 \centering\includegraphics[width=\textwidth]{correlation.epsi}
\end{minipage}
\caption{The number density and the time development of the
correlation function, which roughly speaking is the same as the
backreaction term. Both are compared to the analytic
approximations, and in general there are very good agreement. The
boxcar smoothing, was done to emphasise the secular evolution by
canceling out the oscillations.} \label{fignumberdensity}
\end{figure}

Comparison between the analytical approximations and numerical
computation of the time evolution of the two point correlator
$\langle\varphi^2\rangle$ has been made with excellent agreement
(see figure~\ref{fignumberdensity}). Notice in figure~\ref{fignumberdensity}
the very steep decline in the correlation
function just at the beginning. We interpret it as the first short
time span, where the particles are relativistic. If we look at
the approximation equation~(\ref{eq:psisquared}) for the correlation function
it is singular in the beginning, because we, in making the
approximation, have neglected the $k$-dependence of the energy
$\omega_k$. Obviously we have to take that into account when
$H_0\Delta t \ll 1$. We then find
\begin{equation}\label{eq:relpsisquared}
\langle\varphi^2\rangle_{REL}\simeq\frac{1}{2\pi^2a^2}\int \rmd kk|\beta_k|^2
        =\frac{2n_0}{\sqrt{g|\dot{\phi}_0|}}\left(\frac{a}{a_0}\right)^{-2}.
\end{equation}
If we now, inspired by the form of $\omega_k$, make an ansatz for the correlation function of the form
\begin{equation}\label{eq:realpsisquared}
\langle\varphi^2\rangle = \frac{n_0}{g|\dot{\phi}_0|\sqrt{\Delta t^2 - \Delta t^2_c}}\left(\frac{a}{a_0}\right)^{-3},
\end{equation}
we find, by comparison with equations~(\ref{eq:psisquared}) and (\ref{eq:relpsisquared}) and
taking the appropriate limits, that
\begin{equation}
H_0 \Delta t_c = \frac{H_0}{2\sqrt{g|\dot{\phi}_0|}}.
\end{equation}
Equation (\ref{eq:realpsisquared}) is plotted together with the numerical result in
figure~\ref{fignumberdensity}.
We can interpret $H_0\Delta t_c$ as the number of e-folds where on average the energy of the produced particles is
stored in relativistic modes. Because of the CMBR limit on the quadrupole, 
and recalling the slow-roll result, in
{\it any} model of single field inflation we will have $H_0\Delta t_c\sim 10^{-3}g^{-1/2}$.

It is important to confirm the findings of \cite{Chung:1999a}, that the analytic
approximation for the Bogoliubov coefficients is very precise. The number density
$n_k$ is plotted in figure~\ref{fignumberdensity} and shows excellent agreement.
Notice that at relatively high $k$-values of the wavenumber, compared to the horizon size, the number
density is still significant. The energy density at the moment of particle production per
logarithmic interval scales as $k^3n_k$ and therefore the peak 
energy occurs at values much larger
than $H_0$. This means that the process by far and large is local
in nature and the impact from the expanding background is small.
It assures us that even though we have neglected all the metric
backreaction terms, we still get a result which at least qualitatively agrees
with what a proper treatment would yield.  

In general our results for the background parameters such as the inflaton velocity
and the Hubble parameter are very much in line with \cite{Chung:1999a} and we will
not repeat their analysis here. The reason is that the detailed evolution of the
inflaton $\phi$ or the Hubble parameter $H$ do not have a direct observable impact.
They would only be needed if we wanted to do a detailed analytical modelling.

\subsection{The curvature perturbation}
\begin{figure}[!htbp]
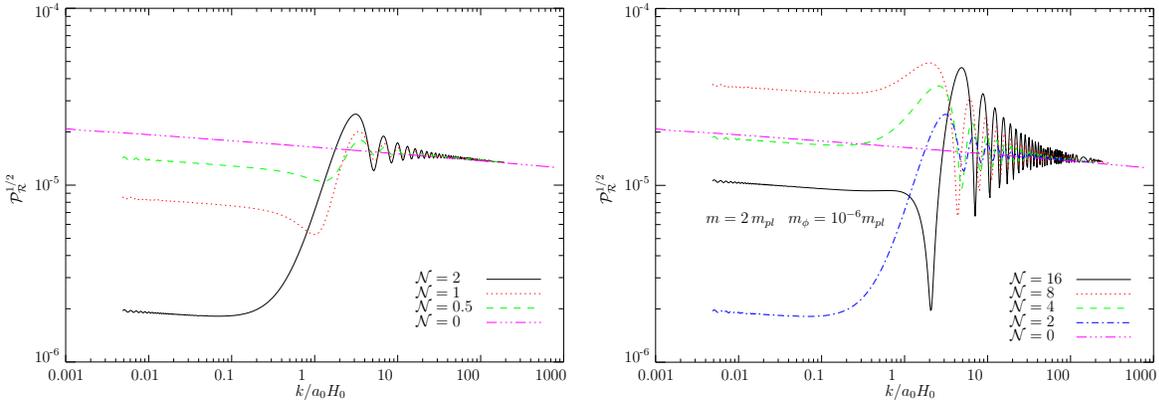

\centering
\begin{minipage}[c]{0.5\textwidth}
 \centering\includegraphics[width=0.95\textwidth]{lowN.epsi}
\end{minipage}%
\begin{minipage}[c]{0.5\textwidth}
 \centering\includegraphics[width=0.95\textwidth]{highN.epsi}
\end{minipage}
\caption{The power spectra from different models. In the right plot the
effective number of degrees of freedom is so high that a perturbative
calculation of the step height cannot be trusted, while the left plot
shows the monotonic evolution of step height for weaker couplings.}
\label{figpowerspectra}
\end{figure}

The essential quantity to consider is the comoving curvature perturbation
${\mathcal R}$. It is found that in the present case there is a qualitative
{\it disagreement} between a numerical evaluation and any analytical prediction,
regardless of whether we use the slow-roll \cite{Abbott:1984a}, the
Stewart--Lyth \cite{Stewart:1993a} or even the semi-numerical method of Leach
{\it et al\/}~\cite{Leach:2001a}. They all disagree with our numerical approach
because they assume that the impact of any change in the fields or fluctuations is
local in the sense that it only affects modes which are just about to cross the
horizon. However, one cannot assume the curvature perturbation to be constant
on super horizon scales if the entropy perturbation
is changing \cite{Wands:2000a,entropy}, which is exactly what happens in our case.
The homogeneous inflaton field
suddenly dumps a large amount of energy into ultra relativistic particles.
It leads to a suppression of
the comoving curvature perturbation on all scales larger than 
the horizon, and the formation
of a step in the power spectrum (see figure~\ref{figpowerspectra}).

On top of the main feature, the step in the potential, there are 
high-frequency oscillations. As is obvious from the graph the
frequency increases with increasing wave number $k$. This suggests a
qualitative interpretation, from a mathematical point of view, of
both the step and the oscillations. The comoving curvature
perturbation ${\mathcal R}$ is related to the inflaton field fluctuation
$Q_k$ according to equation~(\ref{eq:R})
while the power spectrum is defined in the usual way
\begin{equation}
{\mathcal P}_{{\mathcal R}}^{1/2} = \sqrt{\frac{k^3}{2\pi^2}}\left|{\mathcal R}\right|.
\end{equation}
To understand features in ${\mathcal P}_{\mathcal R}$ we can equally well analyse
$Q_k$, because the change in $H$ is negligible and the change in
$\dot{\phi}$ is only transient (see equation~(\ref{eq:R})).

While inside the horizon a single perturbation with definite
wave number $k$ is oscillating with a frequency that
is approximately given as $k/a$ (for a massless field). At the
moment of particle production we can schematically write the
equation for this mode $Q$ as
\begin{equation}
\ddot{Q} + 3H\dot{Q} + \left( \frac{k^2}{a^2} + B(t) \right) Q = 0   ,
\end{equation}
where $B(t)$ is the backreaction term. Now depending on the phase
of $Q$ at the exact moment of particle production the final
amplitude will either be increased or decreased \cite{Adams:2001a}
(see second plot in figure~\ref{figpowerspectra}). Inside the
horizon the phase is highly oscillating $Q\sim \exp(-\rmi k/a t)$ and
we get wiggles on the UV part of the spectrum. Modes which are far
outside the horizon have approximately $k/a = 0$ and the amplitude
changes very slowly, therefore they all experience the same kick
from $B(t)$ making the step in the potential.

In the short interval, where $B^{1/2}(t)$ is greater than the Hubble
mass $3/2 H$, the super-horizon modes begin to oscillate again. In
the perturbative regime this `oscillation' lasts for such a 
short time that the modes do not even cross zero a single time, and we
can compute the impact using simple perturbation theory. On the
other hand, if $B(t)$ is large enough, zero-crossing occurs and
effectively the impact from $B(t)$ is larger than the starting
amplitude and perturbative methods become meaningless.

Let us define the effective step height $N_{\rm eff}$ as the quotient
between the ultraviolet and the infrared part of the power
spectrum ${\mathcal P}_{\mathcal R}^{1/2}$, which is the main observational signature. For
small values of the coupling it is possible to make an analytic
approximation for $N_{\rm eff}$. This is very valuable, because it
enables us to connect the inner parameters of the theory, i.e.~the
coupling, the effective degrees of freedom etc, to the observable:
the effective step height $N_{\rm eff}$.

For low $k$-values, until the particle production starts, $Q_k$
is evolving very slowly and can be taken to be constant. 
Then, suddenly, particle production commences. The main term in the
equation of motion for $Q_k$ equation~(\ref{eq:evolQ2}) is the two point
correlation function. Define the rescaled mode $\tilde{Q} =
a^{3/2}Q$ and use e-folds $N\equiv H\Delta t$ as time variable, then approximately
\begin{equation}\label{eq:rescaledeqom}
\frac{\rmd^2\tilde{Q}}{\rmd N^2} + \left(g^2{\mathcal N}
\frac{\left<\varphi^2\right>}{H^2} - \frac{9}{4}\right)\tilde{Q}
= 0,
\end{equation}
and the boundary conditions are, still for low $k$-values only,
\begin{equation}
    \tilde{Q}(0) = \tilde{Q}_0 ,\tqs
    \frac{\rmd\tilde{Q}}{\rmd N}(0) = \frac{3}{2}\tilde{Q}_0.
\end{equation}
Unfortunately, inserting the expression equation~(\ref{eq:realpsisquared}) for the
correlation function, we get an equation without analytical
solutions. But we can exploit the fact that the
integrated impact of the correlation function is supposed to be
small. Then the exact functional form is not so important, only the
integrated size, and instead of the correlation function can we
substitute an exponential $A \e^{-3N}$ with the correct late time
behaviour. The equation can now be solved in terms of Bessel
functions and plugging in the boundary conditions we get the late
time behaviour and can evaluate the effective step height
\begin{equation}\label{eq:stepsizeI}
N_{\rm eff} = 1 - \frac{1}{9}A + \frac{1}{324}A^2 + \ldots .
\end{equation}
$A$ can be determined by the requirement that the integrated size
of the perturbation has to be the same as for the correlation
function:
\begin{eqnarray} \label{eq:stepsizeII}
\int^\infty_0A\e^{-3N}\rmd N =  g^2{\mathcal N}\int^\infty_0\rmd N\frac{\left<\varphi^2\right>}{H^2}\\ \nonumber
A   =  3g^{5/2}{\mathcal N}\frac{\sqrt{|\dot{\phi}_0|}}{(2\pi)^3H_0}
        \left[\ln\left(\frac{2}{3H_0\Delta t_c}\right)-\gamma_e+3H_0\Delta t_c\right] ,
\end{eqnarray}
where $\gamma_e\simeq 0.577$ is Euler's constant and we have
evaluated the integral to first order in $H_0\Delta t_c$. The result is in
agreement with the numerics.

\section{Models with a changing EOS}
Clearly the last section shows that the impact on the curvature perturbation
on large scales is a generic feature. It has to do with the rapid change
of the potential energy or, alternatively, of the equation of state.
We find our result in agreement with  analyses of similar models,
all posessing the property of having a sudden change in the EOS.

Starobinsky \cite{Starobinsky:1992a} constructed a phenomenological model, 
where the potential has a break in its slope. Roberts {\it et~al\/} \cite{Roberts:1995a}
considered the so-called false vacuum model with quartic potential. This model
is characterized by two periods, where first the potential energy is dominated
by a quartic term in the inflaton, until at a certain moment a cosmological constant 
term takes over the leading role. In the transition period, there
can be a short suspension of inflation, with related change in the EOS.

The index of EOS $\gamma$ is defined as
\begin{equation}
\gamma \equiv \frac{P}{\rho} = \frac{\dot{\phi}^2/2 - V}{\dot{\phi}^2/2 + V},
\end{equation}
where we for simplicity have assumed a single-field model. The evolution equation for the $\phi$-field 
is a second-order equation, and it is clear that a 
change in the EOS can only be sourced by a change in the potential energy. 
In all cases $V_\phi$ diminishes. Because of the equation of motion
\begin{equation}
\ddot{\phi} + 3H\dot{\phi} + V_\phi=0,
\end{equation}
this leads to a large second-order derivative $\ddot{\phi}$ and the inflaton enters a stage
of fast-roll, where approximately
\begin{equation}
\ddot{\phi} + 3H\dot{\phi} \approx 0.
\end{equation}
After some time the balance between the friction term and the source term $V_\phi$ is
reestablished. Looking at equations~(\ref{eq:evolQ}) and (\ref{eq:rescaledeqom}) we can estimate
that a significant change in power spectrum will only happen when
\begin{equation}
\frac{4V_{\phi\phi}}{9H^2}\ge 1,
\end{equation}
for at least one e-fold. Supposing that $V_\phi$ during
one e-fold is diminished by a factor $\epsilon$, the average change during
this period is
\begin{equation}
\frac{\rmd V_\phi}{\rmd N} = -(1-\epsilon)V_\phi = 3(1-\epsilon)H\dot{\phi}.
\end{equation}
Using the last two equations we find a bound on $\epsilon$
\begin{equation}
\epsilon \le \case{1}{4}.
\end{equation}
It is clear that these are very rough estimates, but it does show that to get a
significant change in the power spectrum of the fluctuations a rather large
change in $V_{\phi}$ is required. This could either be due to  interactions with other fields, as in our case, or due to the internal dynamics of the model, as
in the two mentioned cases.

\section{Observational data}

\subsection{Cosmic microwave background}

The CMB temperature
fluctuations are conveniently described in terms of the
spherical harmonics power spectrum
\begin{equation}
C_l \equiv \langle |a_{lm}|^2 \rangle,
\end{equation}
where
\begin{equation}
\frac{\Delta T}{T} (\theta,\phi) = \sum_{lm} a_{lm}Y_{lm}(\theta,\phi).
\end{equation}
Since Thomson scattering polarizes light, there are additional power spectra
coming from the polarization anisotropies. The polarization can be
divided into a curl-free $(E)$ and a curl $(B)$ component, yielding
four independent power spectra: $C_{T,l}$, $C_{E,l}$, $C_{B,l}$ and 
the temperature $E$-polarization cross-correlation $C_{TE,l}$.

The WMAP experiment has reported data only on $C_{T,l}$ and $C_{TE,l}$,
as described in \cite{map1}--\cite{map5}

We have performed the likelihood analysis using the prescription
given by the WMAP collaboration which includes the correlation
between different $C_l$s \cite{map1}--\cite{map5}. Foreground contamination has
already been subtracted from their published data.

In parts of the data analysis we also add other CMB data from
the compilation by Wang {\it et al\/}~\cite{wang3}
which includes data at high $l$.
Altogether this data set has 28 data points.

\subsection{Large scale structure}

The 2dF Galaxy Redshift Survey (2dFGRS) \cite{colless} 
has measured the redshifts 
of more than 230\,000 galaxies with a median redshift of 
$z_{\rm m} \approx 0.11$.  One of the main goals of the survey was to 
measure the galaxy power spectrum on scales up to a few hundred Mpc, 
thus filling in the gap between the small scales covered by earlier 
galaxy surveys and the largest scales where the power spectrum is constrained 
by observations of the CMB.  A sample of the size of the 2dFGRS survey allows 
large-scale structure statistics to be measured with very small random errors.
An initial estimate of the convolved, redshift-space power spectrum of the 
2dFGRS has been determined \cite{percival} for a sample of 160\,000 redshifts. 
On scales $0.02 < k < 0.15h~{\rm Mpc}^{-1}$ the data are robust and the 
shape of the power spectrum is not affected by redshift-space or nonlinear 
effects, though the amplitude is increased by redshift-space distortions.  
A potential complication is the fact that the galaxy power spectrum 
may be biased with respect to the matter power spectrum, i.e.~light does not 
trace mass exactly at all scales.  This is often parametrized by introducing 
a bias factor 
\begin{equation}
b^2(k)\equiv \frac{P_{\rm g}(k)}{P_{\rm m}(k)}, 
\label{eq:biasdef}
\end{equation}
where $P_{\rm g}(k)$ is the power spectrum of the galaxies, and $P_{\rm m}(k)$ 
is the matter power spectrum.  
Indeed it is well established that on scales less than $\sim 10~{\rm Mpc}$ 
different galaxy populations exhibit   different clustering amplitudes, the 
so-called morphology--density relation (see e.g.~\cite{dressler}--\cite{zehavi}).  
Hierarchical merging scenarios also suggest a more complicated picture of biasing 
as it could be non-linear, scale-dependent and stochastic 
\cite{mowhite}--\cite{somerville}.  
However, analysis of the semi-analytic galaxy formation models in 
\cite{berlind}, as well as the simulations in \cite{blanton} suggest that 
the biasing is simple and scale-independent on large scales $k > 0.15h~{\rm Mpc}^{-1}$ 
where the power spectrum is well described by linear theory, and we 
restrict our analysis of the  2dFGRS power spectrum to these scales.   
Two different analyses have demonstrated  
that the 2dFGRS power spectrum is consistent with linear, scale-independent bias \cite{lahav,verde}.   
Thus, the shape of the galaxy power spectrum 
can be used straightforwardly to constrain the shape of the matter power 
spectrum on large scales.

\begin{figure}
\begin{center}
\includegraphics[width=110mm]{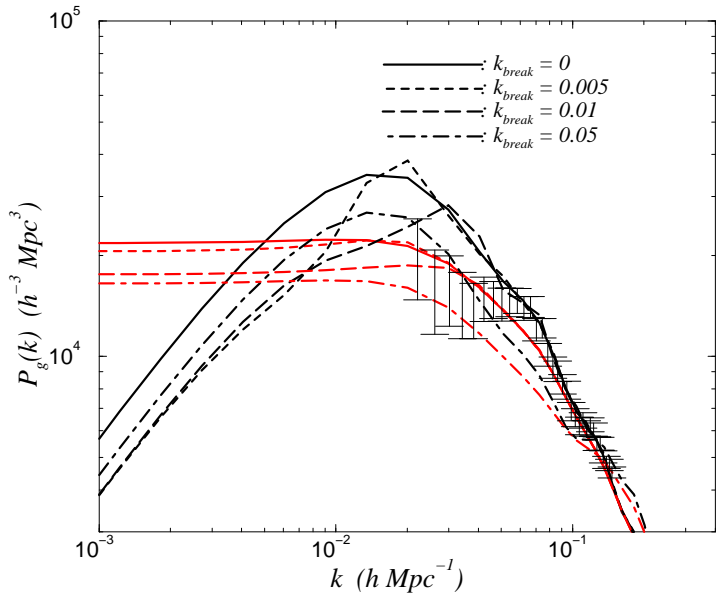}
\caption{The effect of convolving matter power spectra with the 2dFGRS window 
function. The black lines are the matter power spectra computed with CMBFAST for 
four different values of $k_{\rm break}$, the red lines are the corresponding 
spectra after convolution with the window function.  The vertical bars are the 
2dFGRS power spectrum data points.}
\label{fig:convolved}
\end{center}
\end{figure}

\looseness=-1
When looking for steps or other features in the primordial power spectrum 
using the 2dFGRS, one should bear in mind that what is measured is the 
convolution of the true galaxy power spectrum with the 2dFGRS window function 
\cite{percival}, 
\begin{equation}
P_{\rm conv}({\bi k}) \propto \int P_{\rm g}({\bi k}-{\bi q})
|W_k({\bi q})|^2 \rmd^3q,
\label{eq:convolve1}
\end{equation}
where $W$ is the window function.  
In an earlier, simplified attempt to look for steps 
in the primordial power spectrum \cite{elgaroy} it was found that 
the window function of the 2dFGRS more or less washed out any features in 
the primordial power spectrum for comoving momenta $k < 0.1h~{\rm Mpc}^{-1}$. 
The effect on the class of models considered in the present paper 
is illustrated in figure~\ref{fig:convolved} for four different values 
of $k_{\rm break}$.  
However, combining the 2dFGRS power spectrum with CMB data breaks parameter 
degeneracies that are present if each data set is analysed separately, and 
therefore a combination of large-scale structure and CMB data gives 
tighter constraints on the primordial power spectrum than the CMB alone. 

\section{Likelihood analysis}

For calculating the theoretical CMB and matter power spectra we use
the publicly available CMBFAST package \cite{CMBFAST}.
As the set of cosmological parameters we choose
$\Omega_m$, the matter density,
the curvature parameter, $\Omega_b$, the baryon density, $H_0$, the
Hubble parameter, $n_s$, the scalar spectral index of the primordial
fluctuation spectrum, $\tau$, the optical depth to reionization,
$Q$, the normalization of the CMB power spectrum, $b$, the 
bias parameter, and finally the two parameters related to the
step height, $N_{\rm eff}$, and location, $k_{\rm break}$.
We restrict the analysis to geometrically flat models
$\Omega = \Omega_m + \Omega_\Lambda = 1$.

\Table{\label{tab1}The different priors on parameters other than $N_{\rm eff}$
and $k_{\rm break}$ used in the likelihood analysis.}
\br
Parameter & Prior\cr
\mr
$\Omega_m$ & $0.28 \pm 0.14$ (Gaussian) \cr
$h$ & $0.72 \pm 0.08$ (Gaussian) \cr
$\Omega_b h^2$ & 0.014--0.040 (top hat) \cr
$n$ & 0.6--1.4 (top hat) \cr
$\tau$ & 0--1 (top hat) \cr
$Q$ & free \cr
$b$ & free \cr
\br
\endtab

In principle 
one might include even more parameters in the analysis, such as
$r$, the tensor to scalar ratio of primordial fluctuations. However, $r$
is most likely so close to zero that only future high precision 
experiments may be able to measure it. The same is true for other 
additional parameters. Small deviations from slow-roll during inflation
can show up as a logarithmic correction to a simple power-law
spectrum \cite{Hannestad:2000tj}--\cite{griffiths},
or additional relativistic energy
density 
\cite{Jungman:1995bz}--\cite{Dolgov:2002ab}
could be present. However, there is no evidence of any such effect in
the present data and therefore we restrict the analysis to the
`minimal' standard cosmological model.

In this full numerical likelihood analysis we use the free parameters
discussed above with certain priors (see table~\ref{tab1}) determined from cosmological
observations other than CMB and LSS.
In flat models the matter density is restricted by observations
of Type Ia supernovae to be $\Omega_m = 0.28 \pm 0.14$ 
\cite{Perlmutter:1998np}.
The current estimated range for $\Omega_b h^2$ from BBN is 
$\Omega_b h^2 = 0.020 \pm 0.002$ \cite{Burles:2000zk}, 
and finally the HST Hubble key
project has obtained a constraint on 
$H_0$ of $72 \pm 8~{\rm km}~{\rm s}^{-1}~{\rm Mpc}^{-1}$ \cite{freedman}.
The actual maginalization over parameters
other than $N_{\rm eff}$ and $k_{\rm break}$ 
was performed using a simulated annealing procedure~\cite{Hannestad:wx}.

\begin{figure*}[t]
\vspace*{-0.0cm}
\begin{center}
\hspace*{1.0cm}\epsfysize=14truecm\epsfbox{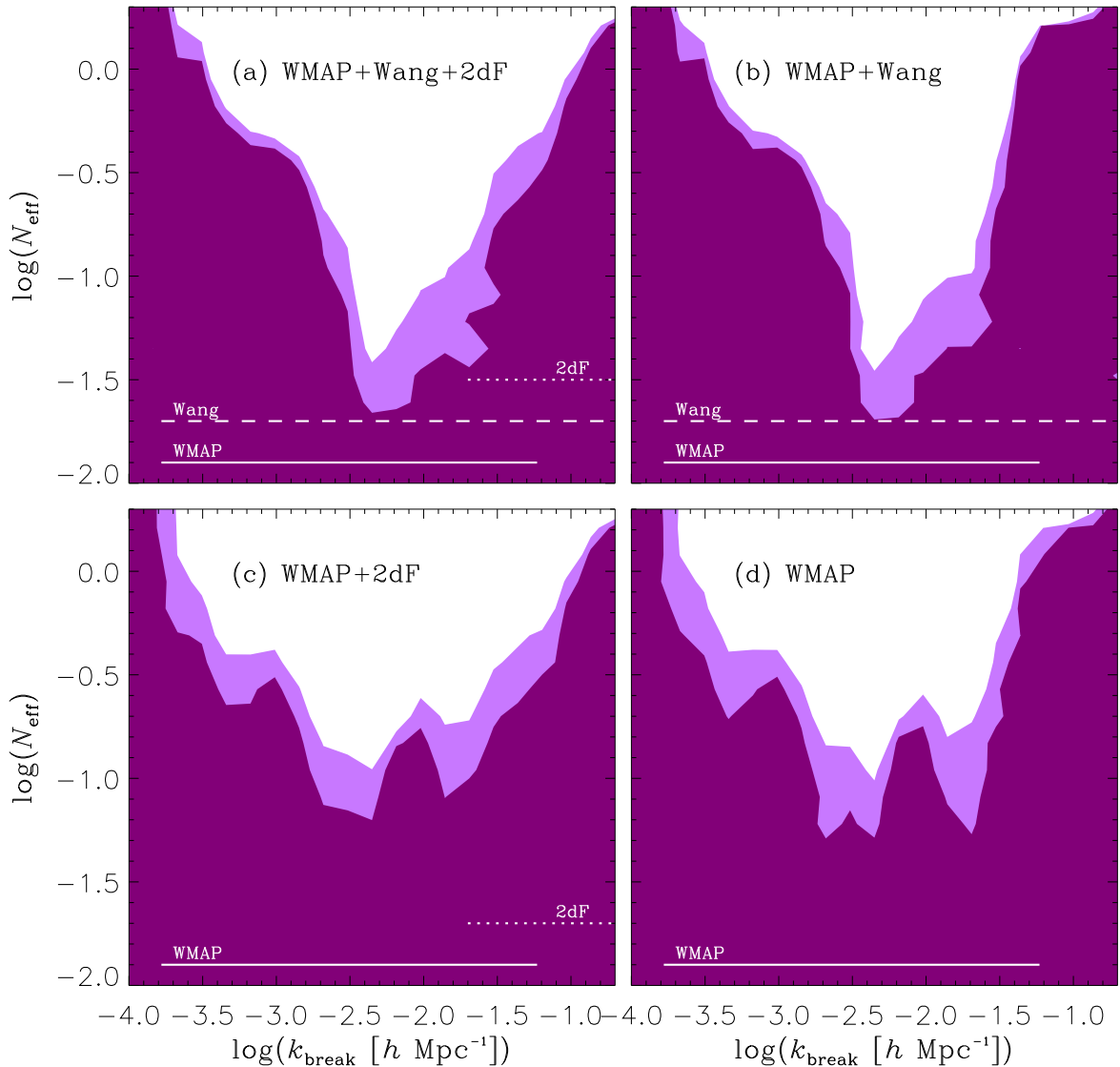}
\end{center}
\vspace*{1cm}
\caption{68\% and 95\% confidence exclusion plot of the parameters
$N_{\rm eff}$ and $k_{\rm break}$ for the four different cases
described in the text. The horizontal lines show the range in $k$-space
covered by various data sets. CMB data have been converted from 
$k$ to $l$ using the approximate prescription
$l \simeq 2 k/H_0$.}
\label{fig:likelihood}
\end{figure*}

In figure~\ref{fig:likelihood} we show results of the likelihood
calculation for various different data sets. In panel (a) results
are shown for an analysis with all available data, WMAP, the Wang
{\it et al\/} compilation and the 2dFGRS power spectrum. 
In panel (b) we show
results for WMAP + Wang, in panel (c) for WMAP + 2dFGRS, and finally
in panel (d) for WMAP data only.
The figure shows the 68\% and 95\% exclusion limits for the parameters
$N_{\rm eff}$ and $k_{\rm break}$. These two contours correspond to
$\Delta \chi^2 = 2.31$ and $\Delta \chi^2 = 6.17$ respectively.
Best fit values and number of degrees of freedom for each of the
four cases are listed in table~\ref{tab2}.

Several things can be noted from this figure. First, the 2dFGRS power 
spectrum only
constrains a step-like feature on scales smaller than $k \sim 0.05 - 0.1h~{\rm Mpc}^{-1}$ 
because the  window function of the survey 
washes out any features on scales larger than~this.

Second, the CMB data are mainly sensitive to scales corresponding 
to $l \sim 50$--100, and only to a lesser extent to features at high $l$.
The main reason for this is that CMB data do not directly probe the
primordial power spectrum, but rather the primordial spectrum convolved
with an effective window function. This will be discussed in detail
in the next subsection. The result of the convolution is that a sharp
step in $P(k)$ appears as a gradual increase in power over a range in $l$ 
in the CMB spectrum. It is relatively easy to mask this effect by changing
other cosmological parameters.
However, scales at $l \lesssim 100$ were outside the particle horizon
at recombination and therefore much less sensitive to changes in most
of the cosmological parameters. This is the reason why the likelihood 
analysis shows a very stringent bound in this range, and not in the
region around the first acoustic peak, as could perhaps have been
expected.

\Table{\label{tab2}Best fit $\chi^2$ and number of degrees of freedom for
each of the cases shown in figure~\ref{fig:likelihood}.}
\br
Case & $\chi^2$ & d.o.f. & $\chi^2$/d.o.f. \cr
\mr
(a) & 1466.9 & 1388 & 1.057 \cr
(b) & 1459.2 & 1371 & 1.064 \cr
(c) & 1440.9 & 1360 & 1.059 \cr
(d) & 1432.2 & 1343 & 1.065\cr
\br
\endtab

Finally, adding the Wang {\it et al\/} compilation to the WMAP data
significantly tightens the bound on $N_{\rm eff}$ over most of the range
in $k$-space, particularly of course at the smaller scales where
the WMAP data suffer from finite angular resolution.
However, in some parts of parameter space the constraint actually
becomes slightly worse by adding other CMB data. This is not 
inconsistent, it just means that the Wang {\it et al\/} data actually
favour a small step and therefore push the total likelihood 
towards higher values of~$N_{\rm eff}$.

In the last panel the effect of the final resolution of the WMAP data
becomes apparent beyond $k \sim 0.05h~{\rm Mpc}^{-1}$, corresponding
to $l \sim 450$, exactly around the scale where the error bars are
no longer cosmic variance limited.

Finally, we note that our analysis is restricted to geometrically flat 
models. However, since spatial geometry affects the CMB spectrum
at scales $l \lesssim 100$ it is possible that the accuracy with
which we have constrained $N_{\rm eff}$ on this scale has been slightly
overestimated.

\subsection{The CMB window function}
CMB data do not directly measure the underlying primordial power
spectrum of fluctuations. Rather, they measure the spectrum folded
with a transfer function in the following sense
\begin{equation}
\label{eq:cl}
C_l = \int \frac{\rmd k}{k} P(k) \Delta_l^2 (k),
\end{equation}
where $\Delta_l^2 (k)$ is the transfer function taken at the present,
$\tau = \tau_0$, and $\tau$ is conformal time.
Following the line-of-sight approach pioneered by Seljak and Zaldariagga
\cite{CMBFAST}
this transfer function can be written as
\begin{equation}
\Delta(k) = \int_0^{\tau_0} \rmd\tau S(k,\tau) j_l [k (\tau-\tau_0)],
\end{equation}
where $S$ is a source function, calculated from the Boltzmann equation,
and $j_l(x)$ is a spherical Bessel function.
However,
in order to get a very rough idea about the effective window function
$w_l(k) = \Delta_l^2(k)/k$ of the CMB we approximate $S$ with a constant
to~obtain
\begin{equation}
\label{eq:window}
w_l(k) \propto \cases{\left(\frac{k \tau_0}{l}\right)^l \sim 0 & for $k \tau_0 \lesssim l$ \cr 
\frac{1}{k^3} & for $k \tau_0 \gtrsim l$.} 
\end{equation}

This simple equation shows several things: (a) A feature at some specific
wavenumber 
$k = k_*$ has the greatest impact on the CMB spectrum 
at $l_* \simeq k_* \tau_0$.
For a flat, matter dominated universe, $\tau_0 = 2/H_0$, yielding
$l_* \simeq 2 k_*/H_0$. (b) The CMB window function is quite
broad, and narrow features in $P(k)$ are accordingly difficult to detect.

In the present model the main detectable feature is a step in the
power spectrum, characterized by the amplitude, $N_{\rm eff}$, and
the location, $k_{\rm break}$.
Adding such a step to an otherwise scale-invariant spectrum yields
a primordial power spectrum 
\begin{equation}
P(k) = A + N_{\rm eff} \Theta(k-k_{\rm break}).
\end{equation}

Calculating $C_l$ from equation~(\ref{eq:cl}), using the window function
equation~(\ref{eq:window}) then gives us
\begin{equation}
\label{eq:step}
C_l \propto
\cases{
\frac{A+N_{\rm eff}}{(l/\tau_0)^2} & for $ k_{\rm break} 
\tau_0 \lesssim l$ \\
\frac{A}{(l/\tau_0)^2}+\frac{N_{\rm eff}}{k^2_{\rm break}} & for $
k_{\rm break} \tau_0 \gtrsim l$.}
\end{equation}

Finally, we can define the ratio, $R$, 
between the CMB spectrum with the step,
and one with exactly the same cosmological parameters, but without
the step
\begin{equation}
\label{eq:ratio}
R =
\cases{
1+\frac{N_{\rm eff}}{A} & for $ k_{\rm break} \tau_0 \lesssim l$ \\
1+\frac{N_{\rm eff}}{A}\left(\frac{l}{k_{\rm break} \tau_0}\right)^2 
& for $
k_{\rm break} \tau_0 \gtrsim l$.}
\end{equation}

\begin{figure}[t]
\vspace*{-0.0cm}
\begin{center}
\hspace*{-0.0cm}\epsfysize=9truecm\epsfbox{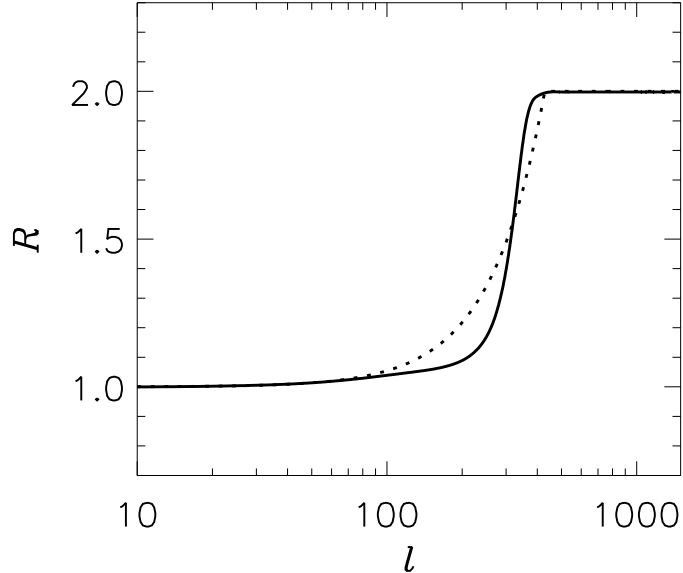}
\end{center}
\vspace*{0cm}
\caption{Ratio of $C_l$ for model with step to model without step.
The full line is the precise numerical calculation and the 
dashed is the approximation in Eq.~(\ref{eq:ratio}).}
\label{fig:ratio}
\end{figure}

In figure~\ref{fig:ratio} we show the ratio of two spectra calculated
with CMBFAST. The first model is a standard $\Omega_m = 1$ CDM model
with $h = 0.7$ and scale-invariant initial spectrum, $P(k) = A$. 
The second model is with the same parameters, but with a step added,
with parameters $N_{\rm eff} = A$ and $k_{\rm break} = 0.05$~Mpc$^{-1}$.
The full line shows the numerical result, and the dashed line shows 
the approximation from equation~(\ref{eq:ratio}). Clearly our very simple
approximation still captures the essential impact of a step-like
feature on the CMB spectrum.
The reason for choosing a CDM model, rather than the observationally
preferred $\Lambda$CDM concordance model, is simply that in the
former case $\tau_0 = 2 H_0^{-1}$, making the analytic expression
in equation~(\ref{eq:ratio}) particularly simple. Using the $\Lambda$CDM
model as the benchmark yields almost exactly the same result.

The conclusion is that
the effective window function $w_l(k)$ for CMB smooths out the
step feature in the underlying primordial power spectrum $P(k)$.
This in turn means that the ability of CMB data to detect step-like
features is significantly degraded. 

A feature which is at first surprising is that the constraint
on the step amplitude $N_{\rm eff}$ is strongest at 
$k_{\rm break} \sim 0.01$, corresponding to $l \sim 90$ which
is significantly above the scale of the first acoustic peak.
The reason is that scales above $l \sim 100$ were outside the
particle horizon at recombination and therefore the CMB spectrum
at these scales is not susceptible to acoustic oscillation effects
which can otherwise mask a step in the primordial spectrum.
At $l \lesssim 30$ the constraint on $N_{\rm eff}$ becomes gradually
weaker simply because cosmic variance increases the error bars on
$C_l$.
Clearly, even though the present CMB data are most sensitive to 
step features at scales corresponding to $l \sim 50$--100 a future
experiment like Planck which is cosmic variance limited out to 
$l \gtrsim 2000$ would be able to put strong constraints on 
$N_{\rm eff}$ at much higher $l$, simply because the greater 
measurement accuracy would break the degeneracy between a step and
changes in other cosmological parameters.

\vspace*{5pt}
\section{Discussion}
\vspace*{3pt}
We have presented a calculation of the perturbation to the power
spectrum that is generated under inflation due to a resonant
coupling of the inflaton to a boson. The calculation has been made 
only under the assumption of slow-roll evolution of the inflaton
before and after the interaction, but no constraints have been put
on the specific form of the potential. In \cite{Chung:1999a} the
same setup was studied, but with a trilinear coupling to a fermion
field. Our results for the perturbation to the primordial power
spectrum are dramatically different to their findings. A step-like
feature is formed, and on top of this, to the ultraviolet side of the
step, there is a highly oscillating transient. For weak couplings
the size of the step scales like $g^{5/2}{\mathcal N}$
and a suppression of power on large scales is observed, while at
larger values of the coupling and the dengeneracy this simple relation
breaks down and suppression or enhancement of the large scale
power relative to the ultraviolet part can occur (see figure~\ref{figpowerspectra}).

The differences we observe with respect to the results of Chung {\it et
al\/} \cite{Chung:1999a} are easily explained from the more limited
analysis they performed. Indeed similar numerical studies support
our conclusion. Easther {\it et
al\/} \cite{Adams:2001a} have investigated
the impact from a step in the potential of the inflaton and found a
very similar feature. They reproduce the oscillations seen in
our model, but no significant amplification is observed. Leach {\it et
al\/} undertook a detailed investigation of the false vacuum model
\cite{Leach:2000a} where they found a suppression of the infrared
part of the spectrum caused by changes in the entropy perturbation
very much like the one we observe. Starobinsky \cite{Starobinsky:1992a} 
explored the very simplest model, where the inflationary potential  has a step,
 and reached the same conclusion. Common for all studies 
is that the effects seen are due to the self contained dynamics of
the field, while in our model they are caused by the mutual
interaction between the inflaton and external fields. This is a
major difference, and while the other models require fine tuning
of the potential, such that the `feature' occurs in the small
range accessible to observational cosmology, our model can be
naturally supported in the context of supersymmetric theories with
extra dimensions or superstring theories. These theories can have
a whole hierarchy of particles with masses comparable to the
Planck mass \cite{Chung:1999a,Lyth:1999a,Malik:2001a}. It would
then be much more natural to expect at least one particle with
parameters, such that the produced feature falls into the range
open to observations \cite{sarkar}. Nonetheless, 
the presented model should be understood as a realistic toy model, 
while a real model should have some basis in particle physics.

\looseness=-1
It would have been desirable to do a full numerical study
including higher order metric terms coming from the 
gravitational interaction of the fields. We think, though, that the overall
features of the phenomena are correctly described in this
analysis. It can be argued, that because the particles are
produced relativistically and the peak wave number is very large
in comparison with the Hubble horizon at that epoch, the process
should be causal in nature, and therefore not much dependent on
the expansion of the universe.

One feature we {\it did} miss in our study, which could be
important, is the quantum treatment of the inflaton fluctuations
and the possibility of cross correlations between the fields. The
step feature is largely due to the backreaction of the coupled
particle directly on the inflaton fluctuations, and therefore the
cross correlation could be almost maximal. This leads to the
possibility, that some of the energy contained in the inflaton
fluctuations oscillates into fluctuations in the other species and
is redshifted away \cite{Bartolo:2001a}. This clearly has to be
addressed in a future study.

Using the recently published WMAP data together with other CMB data 
compiled by Wang {\it et~al\/} and large scale data from the 2DF 
collaboration we have found new upper limits on a break in the 
primordial power spectrum. We did the full likelihood analysis, and our
 conclusions are robust: the derived limits changed only slightly while varying 
the amount of data put into the analysis.
We find that on scales corresponding to $k\sim 0.001$--$0.03h~{\rm Mpc}^{-1}$ the 
$2\sigma$ upper limit on $N_{\rm eff}$---the relative step height---is
$\sim 0.3$. This conclusion is not sensitive to the chosen compilation
of data and can essentially be derived from the WMAP data only. We also note
that the CMB spectrum is mainly sensitive to features at scales corresponding
to $l\sim50$--100 and only to a lesser extent features at high $l$.
This is because scales corresponding to
$l\lesssim 100$ were outside the particle horizon at recombination and 
much less affected by changes in most of the cosmological parameters. 
The constraint is significantly strengthened around $l\sim100$ by adding 
other data than WMAP. This
is indirectly through their tightening of the limits on the cosmological parameters
translating into less freedom to cancel the effect of a break around 
$l \sim 50$--100. While observations of the CMB give us a wonderful tool to 
constrain the cosmology of our universe there is no {\it a priori} reason to 
expect a smooth almost scale invariant power spectrum. The possibility of inherent noise and
deviations in the primordial power spectrum gives us a promising prospect to
probe the earliest history of our universe, but also have the potential of fooling
cosmological parameter estimation at the per cent level. One should remember this
extra uncertainty of `unknown physics', when reading about limits set by
new experiments. While so far there has been no trace of it, the deviation
from full scale invariance and a possible bend in the power spectrum, as has been
inferred by the WMAP team using their data, the 2DF survey and Lyman-alpha forest
data, tells us that we are reaching levels of precision, where features could
begin to crystalize out from the experimental errors.

It should be noted that our likelihood analysis for $k_{\rm break}$
and $N_{\rm eff}$ is in fact quite general and can be applied to other
models which produce a break in the power spectrum. Of course the
oscillating behaviour of the spectrum close to the break is very
specific to the present model, but the width of the window functions
for both CMB and LSS tends to blur the oscillations so that the 
observational constraints are essentially on the step height and position.

Our result for $(k_{\rm break},N_{\rm eff})$ can therefore be compared
for instance to the study of \cite{Bridle:2003sa}.
In that study it was found that the WMAP data show a slight preference
for a step in the power spectrum at $k \simeq 2$--$3 \times 10^{-4}$~Mpc$^{-1}$. 
In their analysis it was assumed that there was no power
above the step scale, which in our language corresponds to 
$N_{\rm eff} \to \infty$.
In fact our WMAP-only analysis shows that the region where
$N_{\rm eff} \to \infty$ and $k_{\rm break} \simeq 1.5$--$2.5 \times 10^{-4}$~Mpc$^{-1}$ 
provides a slightly better fit than no break. However, we find
that this effect is not statistically significant.

Finally we note that our study only probes the scales measurable by
CMB and LSS experiments. On much smaller scales limits on
primordial black holes provide tight constraints on the primordial
power spectrum (see for instance \cite{Bringmann:2001yp})
and can therefore be expected to provide
interesting constraints on step-like features, limiting theories
predicting the occurence of particle production and phase transitions
on a much wider range of scales.

\ack

We acknowledge use of the publicly available CMBFAST package 
written by Uros Seljak and Matthias Zaldarriaga \cite{CMBFAST}
and the use of computing resources at DCSC (Danish Center for
Scientific Computing).

\newcommand\AJ[3]{~Astron. J.{\bf ~#1}, #2~(#3)}
\newcommand\apj[3]{#2 {\it Astrophys.~J.}~{\bf #1} #3}
\newcommand\apjl[3]{~Astrophys. J. Lett. {\bf ~#1}, L#2~(#3)}
\newcommand\ass[3]{~Astrophys. Space Sci.{\bf ~#1}, #2~(#3)}
\newcommand\cqg[3]{~Class. Quant. Grav.{\bf ~#1}, #2~(#3)}
\newcommand\mnras[3]{~Mon. Not. R. Astron. Soc.{\bf ~#1}, #2~(#3)}
\newcommand\mpla[3]{~Mod. Phys. Lett. A{\bf ~#1}, #2~(#3)}
\newcommand\npb[3]{~Nucl. Phys. B{\bf ~#1}, #2~(#3)}
\newcommand\plb[3]{~Phys. Lett. B{\bf ~#1}, #2~(#3)}
\newcommand\prd[3]{#2~{\it Phys.~Rev.~} D {\bf{#1}} #3}
\newcommand\prl[3]{#2~{\it Phys.~Rev.~Lett.~} {\bf{#1}} #3}

\newcommand\prog[3]{~Prog. Theor. Phys.{\bf ~#1}, #2~(#3)}

\end{document}